\documentclass[preprintnumbers,amsmath,amssymb,showpacs]{revtex4}
\usepackage{graphicx}% Include figure files
\usepackage{dcolumn}% Align table columns on decimal point
\usepackage{bm}% bold math

\begin{document}

\title{Detection of genuinely entangled and  non-separable  $n$-partite quantum states}

\author{Ting Gao}
\email{gaoting@hebtu.edu.cn}
\author{Yan Hong}
\affiliation {College of Mathematics and Information Science, Hebei
Normal University, Shijiazhuang 050016, China}

\date{\today}

\begin{abstract}
We investigate  the detection of entanglement in $n$-partite quantum
states.  We obtain practical separability criteria to identify
genuinely entangled and non-separable mixed quantum
 states. No numerical optimization or eigenvalue evaluation is
needed, and our criteria can be evaluated by simple computations
involving components of the density matrix.  We provide examples in
which our criteria perform better than all known separability
criteria. Specifically, we are able to detect genuine
 $n$-partite entanglement which has previously not been identified.  In addition, our criteria can
 be used in today's experiment.
\end{abstract}

\pacs{ 03.65.Ud, 03.67.-a}

\maketitle

Entanglement plays a fundamental role in quantum information
processing and is responsible for many quantum tasks such as quantum
cryptography with Bell's theorem \cite{Ekert91}, quantum dense
coding \cite{BW92}, quantum teleportation
\cite{BBCJPWteleportation93}, quantum communication
\cite{Ekert91,BW92, BBCJPWteleportation93, BBM92, LoChau, YZEPJB05,
GYJPA05} and quantum computation \cite{BD,RB} etc.  Thus,
entanglement is not only the subject of philosophical debates, but
also a new resource for tasks that cannot be performed by means of
classical resources \cite{Bennett98, HorodeckisRMP2009}.

Deciding whether a state is entangled or not has proven to be a very
challenging problem that currently lacks a full computable solution.
 In the bipartite setting, there are some well-known (necessary) criteria for separability, such as
the Bell inequalities \cite{Bell}, positive partial transposition
(PPT) \cite{PeresPPT} (which is also sufficient for two-qubit or one
qubit and one qutrit systems \cite{HorodeckiPPT}), reduction
\cite{HorodeckiReduction, CAprl}, range \cite{HorodeckiRange},
majority \cite{NKmajority}, realignment
\cite{Rrealignment,CWrealignment03, CWrealignment02} and generalized
realignment \cite{ACFgeneralrealignment} etc., which work very well
in many cases, but are far from perfect \cite{HorodeckisRMP2009}.
For multipartite entanglement (more than two parties), the situation
is even more complicated as there exist states that are inseparable
under any fixed partition, but they are still not considered
genuinely multipartite entangled (defined below)
\cite{GuhneNJP2010}. Likewise, there exist states that are
biseparable with respect to each fixed partition, however, they are
not fully separable (for some examples see
Refs.\cite{BennettDiVincenzoMorShor1999PRL,
TothKnappGuhneBriegel2007PRL, AcinBruss2001PRL}).
 Vast areas of multipartite state spaces are still unexplored due to the lack
of suitable tools for detecting and characterizing entanglement.

Recently, G\"{u}hne and Seevinck \cite{GuhneNJP2010} presented a
method for deriving separability criteria within different classes
of 3-qubit and 4-qubit entanglement using density matrix elements.
Huber et al. \cite{MarcusPRL2010} developed a general framework to
identify genuinely multipartite entangled mixed quantum states in
arbitrary-dimensional systems. From the framework introduced in
\cite{MarcusPRL2010}, a k-separability criterion was derived in
\cite{ABMarcus1002.2953}. In addition, we studied the
separability of $n$-partite
 quantum states and  obtained practical separability criteria for different classes of $n$-qubit
and $n$-qudit  quantum states \cite{GaoHong}.

In this paper, we derive novel separability
 criteria to identify genuinely entangled and non-separable $n$-partite mixed quantum
 states. The resulting criteria are easily computable from the density matrix and no
optimization or eigenvalue evaluation is needed.  Below, we first
describe our critera and then provide examples in which we can
detect genuine
 $n$-partite entanglement beyond all previously studied criteria.  Finally, we briefly comment on the ability for our criteria to be implemented in today's experiments without
needing quantum state tomography.

An $n$-partite pure state $|\psi\rangle\in \mathcal{H}_1\otimes
\mathcal{H}_2\otimes\cdots\otimes\mathcal{H}_n$ (dim
$\mathcal{H}=d_i\geq 2$) is called biseparable if there is a
bipartition $j_1j_2\cdots j_k|j_{k+1}\cdots j_n$ such that
\begin{equation}\label{}
|\psi\rangle=|\psi_1\rangle_{j_1j_2\cdots
j_k}|\psi_2\rangle_{j_{k+1}\cdots j_n},
\end{equation}
where $|\psi_1\rangle_{j_1j_2\cdots j_k}$ is the state of particles
$j_1,j_2,\cdots, j_k$, $|\psi_2\rangle_{j_{k+1}\cdots j_n}$ is the
state of particles $j_{k+1},\cdots, j_n$, and  $\{j_1,j_2,\cdots,
j_n \}=\{1,2,\cdots, n\}$. An $n$-partite mixed state $\rho$ is
biseparable if it can be written as a convex combination of
biseparable pure states
\begin{equation}\label{}
 \rho=\sum\limits_{i}p_i|\psi_i\rangle \langle\psi_i|,
\end{equation}
where $|\psi_i\rangle$ might be biseparable under different
partitions.  If an $n$-partite state is not biseparable, then it is
called genuinely $n$-partite entangled.
 An $n$-partite pure state is  fully separable if it is of the
form
\begin{equation}\label{}
|\psi\rangle=|\psi\rangle_1|\psi\rangle_2\cdots|\psi\rangle_n,
\end{equation}
and an $n$-partite mixed state is fully separable if it is a mixture
of fully separable pure states
\begin{equation}\label{}
 \rho=\sum_i p_i |\psi_i\rangle \langle\psi_i|,
\end{equation}
where the $p_i$ forms  a probability distribution, and
$|\psi_i\rangle$ is fully separable. If an $n$-partite state is not
fully separable, then we call it non-separable.
  We  consider
separability criteria of biseparable and fully separable $n$-qubit
and $n$-qudit states.

Throughout this paper, let $\rho$ be a density matrix describing an
$n$-particle system whose state space is Hilbert space
$\mathcal{H}_1\otimes \mathcal{H}_2\otimes\cdots \mathcal{H}_n$,
where dim$\mathcal{H}_l=d_l$, $l=1,2,\cdots, n$. We denote its
entries by $\rho_{i,j}$, where $1\leq i,j\leq d_1d_2\cdots d_n$. We
introduce the further notation of
$|\Phi_{ij}\rangle=|\phi_i\rangle|\phi_j\rangle$ with
$|\phi_i\rangle=|x\cdots xyx\cdots x\rangle\in \mathcal{H}_1\otimes
\mathcal{H}_2\otimes\cdots\otimes\mathcal{H}_n$, where the local
state of $\mathcal{H}_k$ is $|x\rangle$ for $k\not=i$ and
$|y\rangle$ for $k=i$.  Furthermore, let $P$ denote the operator
that performs a simultaneous local permutation on all subsystems in
$(\mathcal{H}_1\otimes
\mathcal{H}_2\otimes\cdots\otimes\mathcal{H}_n)^{\otimes 2}$, while
$P_i$ just performs a permutation on $\mathcal{H}_i^{\otimes 2}$ and
leaves all other subsystems unchanged.

\textbf{Theorem 1} ~ Let $\rho$ be a biseparable $n$-partite density
matrix acting on Hilbert space $\mathcal{H}_1\otimes
\mathcal{H}_2\otimes\cdots\otimes\mathcal{H}_n$, where
dim$\mathcal{H}_l=d_l$, $l=1,2,\cdots, n$. Then
\begin{equation}\label{biseparable-phi}
\sum\limits_{i\neq j}\sqrt{\langle\Phi_{ij}|\rho^{\otimes
2}P|\Phi_{ij}\rangle}\leq\sum\limits_{i\neq
j}\sqrt{\langle\Phi_{ij}|P_i^+\rho^{\otimes
2}P_i|\Phi_{ij}\rangle}+(n-2)\sum\limits_{i}\sqrt{\langle\Phi_{ii}|P_i^+\rho^{\otimes
2}P_i|\Phi_{ii}\rangle},
\end{equation}

If an $n$-partite state $\rho$ does not satisfy the inequality
above, then $\rho$ is genuine $n$-partite entangled.

\textbf{Proof.} ~ To prove that inequality (\ref{biseparable-phi})
is indeed satisfied by all biseparable states $\rho$, let us first
verify that this holds for any pure state $\rho$ which is
biseparable under some partition.

Suppose that $\rho=|\psi\rangle\langle\psi|$ is a
 biseparable pure state under the partition of $\{1,2,\cdots, n\}$
 into two disjoint subsets: $\{1,2,\cdots, n\}=A\cup B$ with $A=\{j_1,j_2,\cdots,
j_k\}$ and $B=\{j_{k+1},\cdots, j_n\}$, and
\begin{equation}\label{}
 \begin{array}{rl}
   |\psi\rangle= & |\psi_1\rangle_{j_1j_2\cdots j_k}|\psi_2\rangle_{j_{k+1}\cdots j_n} \\
    = & (\sum\limits_{i_1,i_2,\cdots, i_k}a_{i_1i_2\cdots i_k}|i_1i_2\cdots i_k\rangle)_{j_1j_2\cdots j_k}
   (\sum\limits_{i_{k+1},\cdots, i_n}b_{i_{k+1}\cdots i_n}|i_{k+1}\cdots i_n\rangle)_{j_{k+1}\cdots j_n} \\
  = & \sum\limits_{i_1,i_2,\cdots, i_n}a_{i_1i_2\cdots i_k}b_{i_{k+1}\cdots i_n}|i_1i_2\cdots
  i_n\rangle_{j_1j_2\cdots j_n},
 \end{array}
\end{equation}
then
\begin{equation}\label{}
\rho_{\sum\limits_{l=1}^{n}i_ld_{j_l+1}d_{j_l+2}\cdots
d_nd_{n+1}+1,\sum\limits_{l=1}^{n}i'_ld_{j_l+1}d_{j_l+2}\cdots
d_nd_{n+1}+1}=a_{i_1i_2\cdots i_k}b_{i_{k+1}\cdots
i_n}a^*_{i'_1i'_2\cdots i'_k}b^*_{i'_{k+1}\cdots i'_n}.
\end{equation}
Here the sum is over all possible values of $i_1,i_2,\cdots, i_n$,
i.e., $\sum_{i_1,i_2,\cdots,
i_n}=\sum_{i_1=0}^{d_{j_1}-1}\sum_{i_2=0}^{d_{j_2}-1}\cdots\sum_{i_n=0}^{d_{j_n}-1}$,
$d_{n+1}=1$.

We will distinguish between the cases in which both indices $i$ and
$j$ corresponding to different, or the same parts $A$ and $B$ in the
bipartition with respect to $|\psi\rangle$.  By calculation, one has
\begin{equation}\label{iandj}
\begin{array}{rl}
   & \sqrt{\langle\Phi_{ij}|\rho^{\otimes
2}P|\Phi_{ij}\rangle}=|\langle\phi_i|\rho|\phi_j\rangle| \\
 = &
 \sqrt{\langle\phi_i|\rho|\phi_i\rangle\langle\phi_j|\rho|\phi_j\rangle}
 \\
 \leq &
 \frac{\langle\phi_i|\rho|\phi_i\rangle+\langle\phi_j|\rho|\phi_j\rangle}{2}
 \\
 = & \frac{\sqrt{\langle\Phi_{ii}|P_i^+\rho^{\otimes
2}P_i|\Phi_{ii}\rangle}+\sqrt{\langle\Phi_{jj}|P_j^+\rho^{\otimes
2}P_j|\Phi_{jj}\rangle}}{2}
\end{array}
\end{equation}
in case of either $i,j\in A$ or $i,j\in B$, and
\begin{equation}\label{iorj}
\begin{array}{rl}
   & \sqrt{\langle\Phi_{ij}|\rho^{\otimes
2}P|\Phi_{ij}\rangle}=|\langle\phi_i|\rho|\phi_j\rangle| \\
 = &
 \sqrt{\langle\phi_0|\rho|\phi_0\rangle\langle\phi_{ij}|\rho|\phi_{ij}\rangle}
 \\
  = & \sqrt{\langle\Phi_{ij}|P_i^+\rho^{\otimes
2}P_i|\Phi_{ij}\rangle}
\end{array}
\end{equation}
in case of either $i\in A,j\in B$ or $i\in B,j\in A$. Here
$|\phi_0\rangle=|xx\cdots x\rangle$, and $|\phi_{ij}\rangle=|x\cdots
xyx\cdots xyx\cdots x\rangle$ such that all particles are in the
state $|x\rangle$ except the $i$th and $j$th particles are in the
state $|y\rangle$. Combining (\ref{iandj}) and (\ref{iorj}) gives
that
\begin{equation}\label{}
\begin{array}{rl}
   & \sum\limits_{i\neq j}\sqrt{\langle\Phi_{ij}|\rho^{\otimes
2}P|\Phi_{ij}\rangle} \\
  = & \sum\limits_{\substack{i\in A,j\in B
\\\textrm{or}~ i\in B,j\in A}}\sqrt{\langle\Phi_{ij}|\rho^{\otimes
2}P|\Phi_{ij}\rangle}+\sum\limits_{\substack{i\neq j ~
\textrm{with}\\ i,j\in A \\\textrm{or}~i,j\in
B}}\sqrt{\langle\Phi_{ij}|\rho^{\otimes
2}P|\Phi_{ij}\rangle} \\
\leq &  \sum\limits_{\substack{i\in A,j\in B
\\\textrm{or}~ i\in B,j\in A}}\sqrt{\langle\Phi_{ij}|P_i^+\rho^{\otimes
2}P_i|\Phi_{ij}\rangle}+\sum\limits_{\substack{i\neq j~
\textrm{with}\\ i,j\in A
\\\textrm{or}~i,j\in
B}}\left(\frac{\sqrt{\langle\Phi_{ii}|P_i^+\rho^{\otimes
2}P_i|\Phi_{ii}\rangle}+\sqrt{\langle\Phi_{jj}|P_j^+\rho^{\otimes
2}P_j|\Phi_{jj}\rangle}}{2}\right) \\
\leq &  \sum\limits_{i\neq
j}\sqrt{\langle\Phi_{ij}|P_i^+\rho^{\otimes
2}P_i|\Phi_{ij}\rangle}+(n-2)\sum\limits_{i}\sqrt{\langle\Phi_{ii}|P_i^+\rho^{\otimes
2}P_i|\Phi_{ii}\rangle}.
\end{array}
\end{equation}
Hence, Ineq.(\ref{biseparable-phi}) is satisfied by all biseparable
$n$-partite pure states.

Next we show that Ineq.(\ref{biseparable-phi}) is also true for all
biseparable $n$-partite mixed states. Indeed, the generalization of
Ineq.(\ref{biseparable-phi}) to mixed states is a direct consequence
of the convexity of its left hand side  and  the concavity of its
right hand side, which we can see in the following.

Suppose that
\begin{equation}\label{}
 \rho=\sum \limits_mp_m\rho_m=\sum
\limits_mp_m|\psi_m\rangle\langle\psi_m|
\end{equation}
is biseparable $n$-partite mixed state, where
$\rho_m=|\psi_m\rangle\langle\psi_m|$ is biseparable. Then,  by
Cauchy-Schwarz inequality
$(\sum\limits_{k=1}^mx_ky_k)^2\leq(\sum\limits_{k=1}^mx_k^2)(\sum\limits_{k=1}^my_k^2)$,
one has
\begin{equation}\label{}
\begin{array}{rl}
 & \sum\limits_{i\neq j}\sqrt{\langle\Phi_{ij}|\rho^{\otimes
2}P|\Phi_{ij}\rangle}\\
\leq & \sum\limits_mp_m\sum\limits_{i\neq
j}\sqrt{\langle\Phi_{ij}|\rho_m^{\otimes 2}
P|\Phi_{ij}\rangle} \\
\leq &  \sum\limits_mp_m\bigg(\sum\limits_{i\neq
j}\sqrt{\langle\Phi_{ij}|P_i^+\rho_m^{\otimes
2}P_i|\Phi_{ij}\rangle}+(n-2)\sum\limits_{i}\sqrt{\langle\Phi_{ii}|P_i^+\rho_m^{\otimes 2}P_i|\Phi_{ii}\rangle}\bigg) \\
= &  \sum\limits_{i\neq
j}\sum\limits_m\sqrt{\langle\phi_0|p_m\rho_m|\phi_0\rangle}\sqrt{\langle\phi_{ij}|p_m\rho_m|\phi_{ij}\rangle}
+(n-2)\sum\limits_{i}\sum\limits_mp_m\langle\phi_i|\rho_m|\phi_i\rangle \\
\leq  &  \sum\limits_{i\neq
j}\sqrt{\sum\limits_m\langle\phi_0|p_m\rho_m|\phi_0\rangle\sum\limits_m\langle\phi_{ij}|p_m\rho_m|\phi_{ij}\rangle}
+(n-2)\sum\limits_{i}\langle\phi_i|\rho|\phi_i\rangle \\
= & \sum\limits_{i\neq j}\sqrt{\langle\Phi_{ij}|P_i^+\rho^{\otimes
2}P_i|\Phi_{ij}\rangle}+(n-2)\sum\limits_{i}\sqrt{\langle\Phi_{ii}|P_i^+\rho^{\otimes
2}P_i|\Phi_{ii}\rangle},
\end{array}
\end{equation}
which finishes the proof of Ineq.(\ref{biseparable-phi}).

It is worth  pointing out that inequality (III) in
Ref.\cite{MarcusPRL2010}, which can be rewritten as
\begin{equation}\label{ineq.(III)MarcusPRL2010}
\sum\limits_{i\neq j}\sqrt{\langle\Phi_{ij}|\rho^{\otimes
2}P|\Phi_{ij}\rangle}\leq   (n-2)\sum\limits_{i,
j}\sqrt{\langle\Phi_{ij}|P_i^+\rho^{\otimes 2}P_i|\Phi_{ij}\rangle},
\end{equation}
 is the corollary of this
theorem. The reason is as follows: Note that the second summation in
inequality (III) of Ref.\cite{MarcusPRL2010}, the right side of
inequality above, can be re-expressed as
\begin{equation}\label{3}
\begin{array}{rl}
(n-2)\sum\limits_{i, j}\sqrt{\langle\Phi_{ij}|P_i^+\rho^{\otimes
2}P_i|\Phi_{ij}\rangle}= & \sum\limits_{i\neq
j}\sqrt{\langle\Phi_{ij}|P_i^+\rho^{\otimes 2}P_i|\Phi_{ij}\rangle}
+(n-2)\sum\limits_{i}\sqrt{\langle\Phi_{ii}|P_i^+\rho^{\otimes
2}P_i|\Phi_{ii}\rangle} \\
 & +(n-3)\sum\limits_{i\neq j}\sqrt{\langle\Phi_{ij}|P_i^+\rho^{\otimes
2}P_i|\Phi_{ij}\rangle}
\end{array}
\end{equation}
in case of $n\geq 3$ and all terms in the third summation term of
the right side of above equality are expectation values of positive
operators, which implies that
\begin{equation}\label{4}
\begin{array}{rl}
 \sum\limits_{i\neq
j}\sqrt{\langle\Phi_{ij}|P_i^+\rho^{\otimes 2}P_i|\Phi_{ij}\rangle}
+(n-2)\sum\limits_{i}\sqrt{\langle\Phi_{ii}|P_i^+\rho^{\otimes
2}P_i|\Phi_{ii}\rangle}\leq & (n-2)\sum\limits_{i,
j}\sqrt{\langle\Phi_{ij}|P_i^+\rho^{\otimes 2}P_i|\Phi_{ij}\rangle}.
\end{array}
\end{equation}
Thus, Ineq.(\ref{ineq.(III)MarcusPRL2010}), inequality (III) in
Ref.\cite{MarcusPRL2010}, follows from Theorem 1 and Ineq.(\ref{4}).

Theorem 1 deserves comments. It is better than inequality (III) of
Ref.\cite{MarcusPRL2010} in the case of genuine multipartite
entanglement detection for $n$-partite quantum states. This
criterion detects genuine $n$-partite entanglement (for $n$-qubit
states such as W state mixed with white noise, and the mixture of
the GHZ state and the W state, dampened by isotropic noise) that had
not been identified so far.

\textit{Example 1} ~ Consider the family of $n$-qubit states
\begin{equation}\label{}
\rho^{(G-W_n)}=\frac{1-\alpha-\beta}{2^n}\mathbb{I}+\alpha|GHZ_n\rangle\langle
GHZ_n|+\beta|W_n\rangle\langle W_n|,
\end{equation}
 the mixture of the GHZ
state and the W state, dampened by isotropic noise.
 Here
\begin{equation}\label{}
|GHZ_n\rangle=\frac{1}{\sqrt{2}}(|00\cdots 0\rangle+|11\cdots
1\rangle)
\end{equation}
 and
\begin{equation}\label{}
|W_n\rangle=\frac{1}{\sqrt{n}}(|00\cdots 001\rangle+|00\cdots
010\rangle+\cdots+|10\cdots 00\rangle)
\end{equation}
 are $n$-qubit GHZ state and
W state, respectively. For this family, our criteria can detect
genuine $n$-partite ($n\geq 4$) entanglement that had not been
identified so far.
 The detection parameter spaces of the
inequality (\ref{biseparable-phi}) in Theorem 1, inequality (III) in
\cite{MarcusPRL2010}, and inequality in \cite{GuhneNJP2010} and
inequality (II) in \cite{MarcusPRL2010} for $n=10$, are illustrated
in  Fig. 1.

\begin{figure}
\begin{center}
{\includegraphics[scale=0.8]{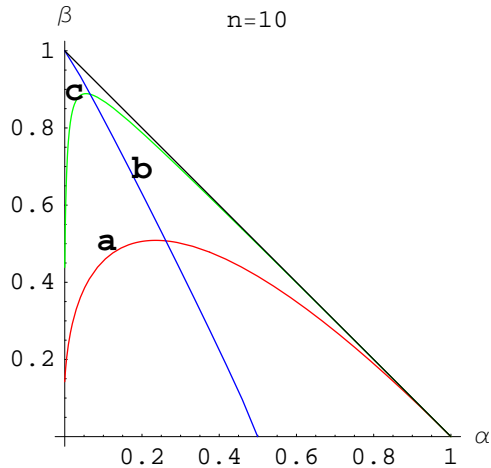}} \caption[Illustration of
the detection parameters of inequality (\ref{biseparable-phi}) in
Theorem 1, inequality in \cite{GuhneNJP2010} and inequality (II) in
\cite{MarcusPRL2010}, and inequality (III) in
\cite{MarcusPRL2010}]{(Color online) Detection quality for the state
$\rho^{(G-W_n)}=\frac{1-\alpha-\beta}{2^n}\mathbb{I}+\alpha|GHZ_n\rangle\langle
GHZ_n|+\beta|W_n\rangle\langle W_n|$, $n=10$. Here the (red) line
$a$ represents the threshold given by inequality
(\ref{biseparable-phi}) in Theorem 1 such that the region above it
identifies genuine 10-partite entanglement.  The regions above lines
$b$ (blue) and $c$ (green) correspond to the genuine entanglement
detected by inequalities (II) in \cite{MarcusPRL2010} (also
\cite{GuhneNJP2010}) and (III) in \cite{MarcusPRL2010} respectively.
The  area enclosed by the red curve $a$, the blue curve $b$, the
green curve $c$, and the $\beta$ axis contains the genuine
10-partite entanglement detected only by inequality
(\ref{biseparable-phi}) in Theorem 1.}
\end{center}
\end{figure}

\textit{Example 2} ~ Let us consider the $n$-qubit state, $W$ states
mixed with white noise,
\begin{equation}\label{}
\rho^{(W_n)}(p)=\frac{p}{2^n}\mathbb{I}+(1-p)|W_n\rangle\langle
W_n|.
\end{equation}
 By Theorem 1 above and
Theorem 3 of Ref.\cite{GaoHong}, we derive that if $0\leq
p<\frac{2^n}{n(2n-3)+2^n}$, then $\rho^{(W_n)}(p)$ is genuine
$n$-partite entangled, while from inequality (III) of
Ref.\cite{MarcusPRL2010}, one can obtain that if $0\leq
p<\frac{2^n}{n^2(n-2)+2^n}$, then $\rho^{(W_n)}(p)$ is genuine
$n$-partite entangled. That is, our criteria detect W state mixed
with white noise, $\rho^{(W_n)}(p)$, for $0\leq
p<\frac{2^n}{n(2n-3)+2^n}$ as genuinely $n$-partite entangled,
whereas inequality (III) of Ref.\cite{MarcusPRL2010} detects it only
for $0\leq p<\frac{2^n}{n^2(n-2)+2^n}$.  For the special case $n=3$
our criteria coincide. When $n=3$, in Ref.\cite{GuhnePhysRep09}
 $\rho^{(W_n)}(p)$  was found to be genuinely multipartite entangled by
 means of the best known entanglement witness up to a threshold of
 $p<\frac{8}{19}$. This bound was then improved to  $p<\frac{8}{17}$
 \cite{GuhneNJP2010, MarcusPRL2010}, which is also our result.
When $n=4$, both Theorem 1 and the previous results
\cite{GuhneNJP2010,GaoHong} detect $\rho^{(W_n)}(p)$ for
$p<\frac{4}{9}\approx 0.444$ as genuine $4$-partite entangled, while
inequality (III) in Ref.\cite{MarcusPRL2010} detects it only for
$p<\frac{1}{3}\approx 0.333$, the fidelity-based witness detects it
only for $p<\frac{4}{15}\approx 0.267$ and the improved witness for
$p<\frac{16}{45}\approx 0.356$ \cite{GuhnePhysRep09}.  However, when
$n=5,6,7,8,9$, Theorem 1
 shows that $\rho^{(W_n)}(p)$ is
genuine multipartite entangled in case of $p<\frac{32}{67}$,
$p<\frac{32}{59}$,
 $p<\frac{128}{205}$, $p<\frac{32}{45}$, $p<\frac{512}{647}$,
respectively, while inequality (III) in Ref.\cite{MarcusPRL2010}
shows that $\rho^{(W_n)}(p)$ is genuine multipartite entangled in
case of
 $p<\frac{32}{107}$, $p<\frac{4}{13}$,
$p<\frac{128}{373}$, $p<\frac{2}{5}$, $p<\frac{512}{1079}$,
respectively. Therefore, our criterion is better than that in
Ref.\cite{MarcusPRL2010}.  W states mixed with white noise,
$\rho^{(W_n)}(p)$, for $\frac{2^n}{n^2(n-2)+2^n}\leq
p<\frac{2^n}{n(2n-3)+2^n}$, as genuine  $n$-partite
($n=5,6,7,8,9,\cdots$) entangled, are for the first time detected by
our criterion. We sum up above results in Table \ref{tab:table1}.
\begin{table}
\caption{\label{tab:table1}The thresholds of the detection for
genuine $n$-partite entanglement for $W$ states mixed with white
noise,
$\rho^{(W_n)}(p)=\frac{p}{2^n}\mathbb{I}+(1-p)|W_n\rangle\langle
W_n|$. The first row represents the number of qubits, while  the
second row and the last row are the thresholds identified by the
inequality (\ref{biseparable-phi}) in Theorem 1 and  the inequality
(III) in \cite{MarcusPRL2010}, respectively. $\rho^{(W_n)}(p)$, for
$\frac{2^n}{n^2(n-2)+2^n}\leq p<\frac{2^n}{n(2n-3)+2^n}$, as genuine
$n$-partite ($n\geq 5$) entangled, are for the first time detected
by inequality (\ref{biseparable-phi}) in Theorem 1.}
\begin{ruledtabular}
\begin{tabular}{cccccccc}
3&4&5&6&7&8&9&$n$\\
\hline
&&&&&&&\\
  $\frac{8}{17}$ & $\frac{4}{9}$ & $\frac{32}{67}$ & $\frac{32}{59}$ & $\frac{128}{205}$ & $\frac{32}{45}$ & $\frac{512}{647}$ & $\frac{2^n}{n(2n-3)+2^n}$ \\
&&&&&&&\\
   $\frac{8}{17}$ & $\frac{1}{3}$ & $\frac{32}{107}$ &
$\frac{4}{13}$ & $\frac{128}{373}$ & $\frac{2}{5}$ &
$\frac{512}{1079}$ &
  $\frac{2^n}{n^2(n-2)+2^n}$
\end{tabular}
\end{ruledtabular}
\end{table}

\textbf{Theorem 2} ~ Every fully separable $n$-partite state $\rho$
satisfies
\begin{equation}\label{fully-separable-1}
\sqrt{\langle\Phi|\rho^{\otimes 2}P|\Phi\rangle}\leq
\bigg(\prod\limits_{A\in S}\langle\Phi|P_A^+\rho^{\otimes
2}P_A|\Phi\rangle\bigg)^{\frac{1}{2^{n+1}-4}}
\end{equation}
for fully separable states $|\Phi\rangle$, where $S$ is the set of
all nonempty proper subsets of $\{1,2,\cdots, n\}$, the permutation
operators $P_A$ are the operators permuting the two copies of all
subsystems contained in  the set $A$, and $P$ is the total
permutation operator, permuting the two copies.

This inequality is equality for fully separable $n$-partite pure
states.

\textbf{Proof.} ~ We start by showing that the
Ineq.(\ref{fully-separable-1}) holds for pure states. So, let us
suppose that $\rho$ is $n$-partite fully separable pure state and
$|\Phi\rangle=|\Phi_1\rangle|\Phi_2\rangle$ with fully separable
$n$-partite states $|\Phi_1\rangle$ and $|\Phi_2\rangle$. The left
side of  Ineq.(\ref{fully-separable-1}) is the absolute value of
matrix element $\langle\Phi_1|\rho|\Phi_2\rangle$:
\begin{equation}\label{}
\sqrt{\langle\Phi|\rho^{\otimes
2}P|\Phi\rangle}=|\langle\Phi_1|\rho|\Phi_2\rangle|,
\end{equation}
since $P$ simply permutes $|\Phi_1\rangle$ and $|\Phi_2\rangle$,
i.e.,
$P|\Phi_1\rangle\otimes|\Phi_2\rangle=|\Phi_2\rangle\otimes|\Phi_1\rangle$.
 Due to
its fully separability, $\rho^{\otimes 2}$ is invariant under
permutation of each element $A$ of $S$:
\begin{equation}\label{}
P_A^+\rho^{\otimes 2}P_A=\rho^{\otimes 2}.
\end{equation}
Thus,
\begin{equation}\label{}
\begin{array}{rl}
   &  \sqrt{\langle\Phi|\rho^{\otimes 2}P|\Phi\rangle}=|\langle\Phi_1|\rho|\Phi_2\rangle| \\
 \leq &
 \sqrt{\langle\Phi_1|\rho|\Phi_1\rangle\langle\Phi_2|\rho|\Phi_2\rangle}=\sqrt{\langle\Phi|\rho^{\otimes 2}|\Phi\rangle} \\
= & \big(\prod\limits_{A\in S}\sqrt{\langle\Phi|\rho^{\otimes
2}|\Phi\rangle}\big)^{\frac{1}{2^n-2}} \\
= & \Big(\prod\limits_{A\in S}\sqrt{\langle\Phi|P_A^+\rho^{\otimes
2}P_A|\Phi\rangle}\Big)^{\frac{1}{2^n-2}},
\end{array}
\end{equation}
as claimed. Here  we have used the positivity of density matrix in
the first inequality and the cardinality $|S|$ of $S$ being $2^n-2$
($S$ has exactly $2^n-2$ elements) in the third equality.
 In fact, for any fully separable pure state $\rho$,  straightforward computation yields
 \begin{equation}\label{}
|\langle\Phi_1|\rho|\Phi_2\rangle| =
 \sqrt{\langle\Phi_1|\rho|\Phi_1\rangle\langle\Phi_2|\rho|\Phi_2\rangle}.
 \end{equation}
Therefore, inequality (\ref{fully-separable-1}) holds with equality
if $\rho$ is fully separable pure state.

It remains to show that Ineq.(\ref{fully-separable-1}) holds if
$\rho$ is mixed state.  Now we suppose that $\rho=\sum p_i\rho_i$ is
a fully separable $n$-partite mixed state, where $\rho_i$ are fully
separable pure states. As the absolute value is convex, i.e.,
$|a+b|\leq|a|+|b|$ for arbitrary complex number $a$ and $b$, and
Ineq.(\ref{fully-separable-1}) is satisfied by fully separable pure
state $\rho_i$, one gets
\begin{equation}\label{1}
\begin{array}{rl}
  & \sqrt{\langle\Phi|\rho^{\otimes 2}P|\Phi\rangle}=|\langle\Phi_1|\rho|\Phi_2\rangle| \\
  \leq & \sum\limits_ip_i|\langle\Phi_1|\rho_i|\Phi_2\rangle|= \sum\limits_ip_i\sqrt{\langle\Phi|\rho_i^{\otimes
  2}|\Phi\rangle} \\
= &  \sum\limits_ip_i\Big(\prod\limits_{A\in
S}\langle\Phi|P_A^+\rho_i^{\otimes
2}P_A|\Phi\rangle\Big)^{\frac{1}{2^{n+1}-4}}.
\end{array}
\end{equation}
By continuously using the H\"{o}lder inequality
\begin{equation}\label{}
\sum\limits_{k=1}^m|x_ky_k|\leq(\sum\limits_{k=1}^m|x_k|^p)^{\frac{1}{p}}(\sum\limits_{k=1}^m|y_k|^q)^{\frac{1}{q}}
~ (p,q>1,\dfrac{1}{p}+\dfrac{1}{q}=1),
\end{equation}
we obtain that
\begin{equation}\label{2}
\begin{array}{rl}
  & \sum\limits_ip_i\Big(\prod\limits_{A\in
S}\langle\Phi|P_A^+\rho_i^{\otimes
2}P_A|\Phi\rangle\Big)^{\frac{1}{2^{n+1}-4}} \\
 \leq &  \Big(\prod\limits_{A\in
S}\langle\Phi|P_A^+(\sum\limits_ip_i^2\rho_i^{\otimes
2})P_A|\Phi\rangle\Big)^{\frac{1}{2^{n+1}-4}} \\
\leq & \Big(\prod\limits_{A\in S}\langle\Phi|P_A^+\rho^{\otimes
2}P_A|\Phi\rangle\Big)^{\frac{1}{2^{n+1}-4}},
\end{array}
\end{equation}
where  in the second inequality we have used $\rho^{\otimes
2}-\sum\limits_ip_i^2\rho_i^{\otimes 2}=\sum\limits_{i\neq
j}p_ip_j\rho_i\otimes\rho_j\geq 0$, since density matrices $\rho_i$
are positive semi-definite, i.e., $\rho_i\geq 0$. Combining
Ineqs.(\ref{1}) and (\ref{2}) gives Ineq.(\ref{fully-separable-1}),
as required. This completes the proof.

In particular, if $\rho$ is fully separable $n$-qubit state, then
this theorem for $|\Phi\rangle=|00\cdots 0\rangle|11\cdots 1\rangle$
implies
\begin{equation}\label{n-qubit-fully-separable}
 |\rho_{1,2^n}|\leq
 \left(\rho_{2,2}\rho_{3,3}\rho_{4,4}\cdots\rho_{2^n-1,2^n-1}\right)^{\frac{1}{2^n-2}},
\end{equation}
the first inequality of Theorem 4 in Ref.\cite{GaoHong}, which is
necessary and sufficient condition \cite{GaoHong} for GHZ state
mixed with white noise,
$\rho(p)=(1-p)|\textrm{GHZ}_n\rangle\langle\textrm{GHZ}_n|+\dfrac{p}{2^n}\textrm{I}$
as fully separable, where
$|\textrm{GHZ}_n\rangle=\frac{1}{\sqrt{2}}(|00\cdots
\rangle+|11\cdots 1\rangle)$.

For  detection of non-separable quantum states, Theorem 2 is as
strong as the PPT criterion and criterion $(\ast)$ in
Ref.\cite{ABMarcus1002.2953}. Consider the most general maximally
entangled state ( general GHZ
 states ) for $n$-qudits mixed with white noise
\begin{equation}\label{max-entangle+white-noise}
\rho=p|\Psi\rangle\langle\Psi|+\frac{1-p}{d^n}\mathbb{I}_{d^n},
\end{equation}
where
\begin{equation}\label{max-entangle-qudit}
|\Psi\rangle=\frac{1}{\sqrt{d}}\sum\limits_{i=0}^{d-1}|i\rangle^{\otimes
n}.
\end{equation}
Direct calculation of inequality (\ref{fully-separable-1}) yields
that these states are  non-separable (not fully separable) if
\begin{equation}\label{}
p>\frac{1}{1+d^{n-1}},
\end{equation}
which is exactly the threshold detected by PPT criterion and
criterion $(\ast)$ in Ref.\cite{ABMarcus1002.2953}.

\textbf{Theorem 3} ~ Suppose that $\rho$ is a fully separable
$n$-partite state. Then the following inequality
\begin{equation}\label{fully-separable-2}
\sum\limits_{i\neq j}\sqrt{\langle\Phi_{ij}|\rho^{\otimes
2}P|\Phi_{ij}\rangle} \leq \sum\limits_{i\neq
j}\sqrt{\langle\Phi_{ij}|P_i^+\rho^{\otimes 2}P_i|\Phi_{ij}\rangle}
\end{equation}
holds with equality if $\rho$ is a pure state.

\textbf{Proof.} ~ Note that  the left side of the inequality
(\ref{fully-separable-2}) minus the right side of
(\ref{fully-separable-2}) is a convex function of the matrix $\rho$
entries (since the left side is the summation of absolute values of
density matrix elements and the right hand is the summation of the
square root of a product of two diagonal density matrix  elements).
Consequently, it suffices to prove the validity for fully separable
pure states and validity for mixed states is guaranteed.

Similar to the proof of Theorem 1 we need only to prove that
Ineq.(\ref{fully-separable-2}) holds for fully separable pure
states.  Suppose that $\rho$ is a pure state. Since $\rho$ is a
fully separable pure state, this gives
\begin{eqnarray}
 |\langle\phi_i|\rho|\phi_j\rangle| &=& \sqrt{\langle\phi_i|\rho|\phi_i\rangle\langle\phi_j|\rho|\phi_j\rangle}
 =\sqrt{\langle\phi_0|\rho|\phi_0\rangle\langle\phi_{ij}|\rho|\phi_{ij}\rangle}, \\
 P_i^+\rho^{\otimes 2}P_i  &=& \rho^{\otimes 2},
\end{eqnarray}
where $|\phi_0\rangle$ and $|\phi_{ij}\rangle$ are the same as that
in Theorem 1. Applying these two equalities, we have
\begin{equation}\label{}
\sum\limits_{i\neq j}\sqrt{\langle\Phi_{ij}|\rho^{\otimes
2}P|\Phi_{ij}\rangle}=\sum\limits_{i\neq j}
|\langle\phi_i|\rho|\phi_j\rangle|=\sum\limits_{i\neq
j}\sqrt{\langle\phi_i|\rho|\phi_i\rangle\langle\phi_j|\rho|\phi_j\rangle}=\sum\limits_{i\neq
j}\sqrt{\langle\phi_0|\rho|\phi_0\rangle\langle\phi_{ij}|\rho|\phi_{ij}\rangle}=\sum\limits_{i\neq
j}\sqrt{\langle\Phi_{ij}|P_i^+\rho^{\otimes 2}P_i|\Phi_{ij}\rangle},
\end{equation}
as desired. This completes the proof.

For the $n$-qubit W state mixed with white noise, $\rho^{(W_n)}(p)$,
equation (\ref{fully-separable-2}) detects entanglement for
\begin{equation}\label{}
p<\frac{2^n}{2^n+n},
\end{equation}
that is, $\rho^{W_n}(p)$ is entangled (not fully separable) if
$p<\frac{2^n}{2^n+n}$.

Our criteria are experimentally accessible without quantum state
tomography. Each term in the left hand side of our criteria can be
determined by measuring two observables, while each term in the
right hand side can be determined by one observable. For any fixed
$|\Phi_{ij}\rangle$, Eq.(\ref{biseparable-phi}) and
Eq.(\ref{fully-separable-2}) can be determined by $n^2+1$ and
$n^2-n+1$ density matrix elements, respectively. For any fixed
$|\Phi\rangle$, Eq.(\ref{fully-separable-1}) can be determined by
$2^n-1$ density matrix elements. Compared to the
$(d^2_1-1)(d^2_2-1)\cdots (d^2_n-1)$ measurements needed for quantum
state tomography, which requires an exponentially increasing since
$(d^2_1-1)(d^2_2-1)\cdots (d^2_n-1)=(d^2-1)^n$ in case of all
subsystems with same dimension $d$, the numbers of density matrix
elements in our criteria not only grows significantly slower with
$n$, but   have great advantage of being independent of the
dimension $d_l$ of the subsystem $l$, $l=1,2,\ldots,n$.

The observables associated with each term (diagonal matrix elements)
of the right hand side in Eq.(\ref{biseparable-phi}) and
Eq.(\ref{fully-separable-2}) can be implemented by means of local
observables, which can be seen from the following expressions
$|\phi_0\rangle\langle\phi_0|=P^{\otimes n}$,
$|\phi_{ij}\rangle\langle\phi_{ij}|=P^{\otimes(i-1)}\otimes Q\otimes
P^{\otimes(j-i-1)}\otimes Q\otimes P^{\otimes(n-j)}$, and
$|\phi_i\rangle\langle\phi_i|=P^{\otimes(i-1)}\otimes Q\otimes
P^{\otimes(n-i)}$, where $P=|x\rangle\langle x|$ and
$Q=|y\rangle\langle y|$. Similarly, each term of the right hand side
in Eq.(\ref{fully-separable-1}) can also be determined by local
measurement. Thus, determining one diagonal matrix element requires
only a single local observable.

From $\sqrt{\langle\Phi_{ij}|\rho^{\otimes
2}P|\Phi_{ij}\rangle}=|\langle\phi_i|\rho|\phi_j\rangle|$ and
$\sqrt{\langle\Phi|\rho^{\otimes
2}P|\Phi\rangle}=|\langle\Phi_1|\rho|\Phi_2\rangle|$, next, we
should determine modulus of the off diagonal elements
$|\langle\phi_i|\rho|\phi_j\rangle|$ by measuring two observables
$O_{ij}$ and $\tilde{O}_{ij}$, and
$|\langle\Phi_1|\rho|\Phi_2\rangle|$ by measuring $O$ and
$\tilde{O}$, since  $\langle
O_{ij}\rangle=2\mathrm{Re}\langle\phi_i|\rho|\phi_j\rangle$,
$\langle
\tilde{O}_{ij}\rangle=-2\mathrm{Im}\langle\Phi_i|\rho|\phi_j\rangle$,
$\langle O\rangle=2\mathrm{Re}\langle\Phi_1|\rho|\Phi_2\rangle$, and
$\langle
\tilde{O}\rangle=-2\mathrm{Im}\langle\Phi_1|\rho|\Phi_2\rangle$.
Here
$O_{ij}=|\phi_i\rangle\langle\phi_j|+|\phi_j\rangle\langle\phi_i|$,
$\tilde{O}_{ij}=-\mathrm{i}|\phi_i\rangle\langle\phi_j|+\mathrm{i}|\phi_j\rangle\langle\phi_i|$,
$O=|\Phi_1\rangle\langle\Phi_2|+|\Phi_2\rangle\langle\Phi_1|$, and
$\tilde{O}=-\mathrm{i}|\Phi_1\rangle\langle\Phi_2|+\mathrm{i}|\Phi_2\rangle\langle\Phi_1|$.

Without loss of generality, let $i<j$. From
\begin{equation}\label{}
\begin{array}{rl}
  O_{ij}= & \frac{1}{2} P^{\otimes(i-1)}\otimes M \otimes P^{\otimes(j-i-1)}\otimes
M\otimes P^{\otimes(n-j)}
\\
& + \frac{1}{2} P^{\otimes(i-1)}\otimes \tilde{M} \otimes
P^{\otimes(j-i-1)}\otimes \tilde{M}\otimes P^{\otimes(n-j)},
\end{array}
\end{equation}
\begin{equation}\label{}
\begin{array}{rl}
  \tilde{O}_{ij} =& \frac{1}{2} P^{\otimes(i-1)}\otimes M \otimes P^{\otimes(j-i-1)}\otimes
\tilde{M}\otimes P^{\otimes(n-j)}
\\
& - \frac{1}{2} P^{\otimes(i-1)}\otimes \tilde{M} \otimes
P^{\otimes(j-i-1)}\otimes M\otimes P^{\otimes(n-j)},
\end{array}
\end{equation}
where $M=|y\rangle\langle x|+|x\rangle\langle y|$,
$\tilde{M}=\mathrm{i}|y\rangle\langle x|-\mathrm{i}|x\rangle\langle
y|$, one can determine the left hand side in Eq.(5) by $2(n^2-n)$
local observables.

  Suppose
$|\Phi_1\rangle=|x_1x_2\ldots x_n\rangle$,
$|\Phi_2\rangle=|y_1y_2\ldots y_n\rangle$. Let
$R_l=|y_l\rangle\langle x_l|+|x_l\rangle\langle y_l|$ and
$\tilde{R}_l=\mathrm{i}|y_l\rangle\langle
x_l|-\mathrm{i}|x_l\rangle\langle y_l|$, $l=1,2,\ldots,n$. Following
the method of \cite{GuhneLu-2007PRL, M.SeevinkJ.Uffink}, element
$\sqrt{\langle\Phi|\rho^{\otimes 2}P|\Phi\rangle}$ can be obtained
from two local measurement settings $R_i$ and $\tilde{R}_i$, given
by
\begin{equation}\label{}
\mathcal{M}_l=\left[\cos\left(\frac{l\pi}{n}\right)
R_l+\sin\left(\frac{l\pi}{n}\right)\tilde{R}_l\right]^{\otimes n}, ~
l=1,2,\ldots,n,
\end{equation}
\begin{equation}\label{}
\mathcal{\tilde{M}}_l=\left[\cos\left(\frac{l\pi+\pi/2}{n}\right)
R_l+\sin\left(\frac{l\pi+\pi/2}{n}\right)\tilde{R}_l\right]^{\otimes
n}, ~ l=1,2,\ldots,n.
\end{equation}
These operators obey
\begin{eqnarray}
\sum_{l=1}^n(-1)^l\mathcal{M}_l &=& nO, \\
\sum_{l=1}^n(-1)^l\mathcal{\tilde{M}}_l &=& n\tilde{O},
\end{eqnarray}
which can be proved in the same way as \cite{GuhneLu-2007PRL,
M.SeevinkJ.Uffink}.

Therefore in total at most $\frac{5(n^2-n)}{2}+n+1$,
$\frac{5(n^2-n)}{2}+1$, and $2^n+2n-2$ local observables are needed
to test our separability criteria Eq.(\ref{biseparable-phi}),
Eq.(\ref{fully-separable-1}), and Eq.(\ref{fully-separable-2}),
respectively.

In conclusion, we investigate  $n$-partite
 quantum states from elements of density matrices and derive practical separability
 criteria to identify genuinely entangled and non-separable $n$-partite mixed quantum
 state. We show cases in which our criteria is
 stronger than all known separability criteria. In fact, our criteria detect genuine
 $n$-partite entanglement that had not been identified so far.
 It has the added appeal of being relatively easy to compute and
 requiring far fewer measurements to implement experimentally
 compared to full quantum tomography.

\vspace{0.6cm}

This work was supported by the National Natural Science Foundation
of China under Grant No: 10971247, Hebei Natural Science Foundation
of China under Grant Nos: F2009000311, A2010000344.

\end{document}